\documentstyle[12pt]{article}
\begin{document}

\title{\textbf{Power-law entropy corrected holographic and new agegraphic $f(R)$-gravity models}}

\author{K. Karami$^{1,2}$\thanks{E-mail: KKarami@uok.ac.ir} ,
M.S. Khaledian${^1}$\thanks{E-mail:
MS.Khaledian@uok.ac.ir}\\$^{1}$\small{Department of Physics,
University of Kurdistan, Pasdaran St., Sanandaj,
Iran}\\$^{2}$\small{Research Institute for Astronomy
$\&$ Astrophysics of Maragha (RIAAM), Maragha, Iran}\\
}

\maketitle

\begin{abstract}
Motivated by the recent works of us \cite{karamikhaled}, we
establish a correspondence between $f(R)$-gravity with the power-law
entropy corrected holographic dark energy (PLECHDE) and new
agegraphic dark energy (PLECNADE) models. We reconstruct
corresponding $f(R)$-gravities and obtain the equation of state
parameters. We show that the selected $f(R)$-gravity models can
accommodate the transition from the quintessence state to the
phantom regime as indicated by the recent observations.
\end{abstract}
\clearpage

\section{The theory of $f(R)$ modified gravity}
The Friedmann equations in $f(R)$-gravity are given by
\cite{NojiriOdin2009}
\begin{equation}
\frac{3}{k^2}H^2= \rho_m+\rho_R,~~~~~~~~~~~~~~~~\label{FiEq1}
\end{equation}
\begin{equation}
\frac{1}{k^2}(2\dot{H}+3H^2)=-(p_m+p_R),\label{FiEq2}
\end{equation}
where
\begin{equation}
\rho_{R} = \frac{1}{k^2}\left[-\frac{1}{2}f(R)+3(\dot{H}+
H^2)f'(R)-18(4 H^2\dot{H}+ H\ddot{H})f''(R)\right],~~~~~~~~~\label{density}
\end{equation}
\begin{equation}
\begin{array}{l}
p_{R}= \frac{1}{k^2}\Big[\frac{1}{2}f(R)-(\dot{H}+3H^2)f'(R)+\nonumber\\
~~~~~~~~~~~~6(8H^2\dot{H}+6H\ddot{H}+4{\dot{H}}^2+\dot{\ddot{H}})f''(R)+36(H\ddot{H}+4H^2\dot{H})^2f'''(R)\Big],\label{pressure}
\end{array}
\end{equation}
with
\begin{equation}
R =6(\dot{H}+2H^2).\label{R}
\end{equation}
Here, $\rho_{m}$ and $p_{m}$ are the energy density and pressure of
the matter, respectively. Also $\rho_{R}$ and $p_{R}$ are due to the
contribution of $f(R)$-gravity.

The equation of state (EoS) parameter of the curvature contribution
is defined as \cite{Nozari}
\begin{eqnarray}
\omega_{R}=\frac{p_R}{\rho_R}=-1+\frac{p_R+\rho_R}{\rho_R}.\label{wHDETotal}
\end{eqnarray}
For a given $a=a(t)$, by the help of Eqs. (\ref{density}) and
(\ref{pressure}) one can reconstruct the $f(R)$-gravity according to
any DE model given by the EoS $p_R=p_R(\rho_R)$ or
$\rho_R=\rho_R(a)$ \cite{NojiriOdin2006}.

Here we consider the two classes of scale factors, the first one is
given by \cite{NojiriOdin2006,Sadjadi}
\begin{equation}
a(t)=a_0(t_s-t)^{-h},~~~t\leq t_s,~~~h>0.\label{a}
\end{equation}
Using Eqs. (\ref{R}) and (\ref{a}) one can obtain
\begin{equation}
H=\frac{h}{t_s-t}=\left[\frac{h}{6(2h+1)}R\right]^{1/2},~~~\dot{H}=H^2/h.\label{respect
to r}
\end{equation}
For the second class of scale factors defined as
\cite{NojiriOdin2006}
\begin{equation}
a(t)=a_0t^h,~~~h>0,\label{aQ}
\end{equation}
one can get
\begin{equation}
H=\frac{h}{t}=\left[\frac{h}{6(2h-1)}R\right]^{1/2},~~~\dot{H}=-H^2/h.\label{respect
to rQ}
\end{equation}
In sections 2 to 4 using the two classes of scale factors (\ref{a})
and (\ref{aQ}), we reconstruct different $f(R)$-gravities according
to the PLECHDE and PLECNADE models.

\section{PLECH $f(R)$-gravity model}

The holographic dark energy (HDE) density is given by \cite{Li6}
\begin{equation}\label{holo den}
    \rho_{\Lambda}=\frac{3M_p^2c^2}{L^2},
\end{equation}
where $c$ and $M_p^{-2}=8\pi G$ are the numerical constant and
reduced Planck mass. Also $L$ is the IR-cutoff of the universe. This
model can be modified due to the power-law corrections to the
entropy which appears in dealing with the entanglment of quantum
fields in and out the horizon \cite{das}. The power-law corrected
entropy takes the form \cite{das}
\begin{equation}\label{entropy}
    S=\frac{A}{4G}\Big[1-K_{\alpha}A^{1-\alpha/2}\Big],
\end{equation}
where $\alpha$ is a dimensionless constant whose value is currently under debate and
\begin{equation}\label{K cons}
    K_{\alpha}=\frac{\alpha(4\pi)^{\alpha/2-1}}{(4-\alpha)r_c^{2-\alpha}}.
\end{equation}
Also $r_c$ is the crossover scale. Using (\ref{entropy}), the
PLECHDE density was introduced as \cite{shay}
\begin{equation}\label{PLECHDE}
    \rho_{\Lambda}=M_p^2\Big(\frac{3c^2}{L^2}-\frac{\beta}{L^{\alpha}}\Big),
\end{equation}
where in the absence of correction terms ($\alpha=\beta=0$) yields
the well-known HDE density \cite{das}.

For a flat universe  we have $L=R_{\rm h}$, where $R_h$ is the event
horizon defined as
\begin{equation}
R_h=a\int_t^{\infty}\frac{{\rm d}t}{a}=a\int_a^{\infty}\frac{{\rm
d}a}{Ha^2}.\label{L0}
\end{equation}
For the first class of scale factors (\ref{a}) and using Eq.
(\ref{respect to r}), the future event horizon $R_h$ yields
\begin{equation}
R_h=a\int_t^{t_s}\frac{{\rm
d}t}{a}=\frac{t_s-t}{h+1}=\frac{1}{h+1}\sqrt{\frac{6h(2h+1)}{R}}.\label{L}
\end{equation}
Here due to having $R_h>0$ then $-1<h$.

Inserting Eq. (\ref{L}) into (\ref{PLECHDE}) one can get
\begin{equation}\label{p}
\rho_{\Lambda}=M_p^2\left[\frac{c^2(h+1)^2}{2h(2h+1)}R-
\frac{\beta(h+1)^\alpha}{\big(6h(2h+1)\big)^{\frac{\alpha}{2}}}R^\frac{\alpha}{2}\right].
\end{equation}
Equating (\ref{density}) with (\ref{p}) gives
\begin{eqnarray}\label{dif eq4}
2R^2 f''(R)-(h+1)R
f'(R)+(2h+1)f(R)~~~~~~~~~~~~~~~~~~~\nonumber\\+\left[\frac{c^2(h+1)^2}{h}R-
\frac{2\beta(2h+1)(h+1)^\alpha}{\big(6h(2h+1)\big)^{\frac{\alpha}{2}}}R^\frac{\alpha}{2}\right]=0.
\end{eqnarray}

Solving Eq. (\ref{dif eq4}) yields the PLECHDE $f(R)$-gravity as
\begin{eqnarray}
f(R)=\lambda_+R^{m_+}+\lambda_-R^{m_-}-\frac{c^2(h+1)^2R}{h^2}
~~~~~~~~~~~~~~~~~\nonumber\\
-\frac{4\beta(2h+1)(h+1)^\alpha R^{\frac{\alpha}{2}}}
{\big(h(\alpha-4)-(\alpha-1)(\alpha-2)\big)\big(6h(2h+1)\big)^{\frac{\alpha}{2}}},\label{fR-PLECHDE}
\end{eqnarray}
where
\begin{equation}\label{alpha}
m_\pm=\frac{3+h\pm\sqrt{h^2-10h+1}}{4}.
\end{equation}
Also $\lambda_\pm$ are the integration constants which can be found
from the following boundary conditions \cite{NojiriOdin1}
\begin{equation}
f(R_0)=-2R_0,\label{bc1}
\end{equation}
\begin{equation}
f'(R_0)\sim0,\label{bc2}
\end{equation}
where $R_0\sim(10^{-33}{\rm eV})^2$ is the current curvature.\\
Applying the above boundary conditions to the solution (\ref{fR-PLECHDE})
one can obtain $\lambda_\pm$ as
\begin{eqnarray}\label{landa}
    \lambda_\pm=\frac{1}{(m_\pm-m_\mp)R_0^{m\pm}}
    \Big[\Big(2m_\mp+\frac{(1-m_\mp)c^2(h+1)^2}{h}\Big)R_0+
    ~~~~~~~~~~~~\nonumber\\
    \frac{2\beta(\alpha-2m_\mp)(2h+1)(h+1)^\alpha R_0^{\frac{\alpha}{2}}}{\big(h(\alpha-4)-
    (\alpha-1)(\alpha-2)\big)\big(6h(2h+1)\big)^{\frac{\alpha}{2}}}\Big].
\end{eqnarray}
Replacing Eq. (\ref{fR-PLECHDE}) into (\ref{wHDETotal}) and using
(\ref{respect to r}) and (\ref{L}) one can get the EoS parameter of
the PLECHDE $f(R)$-gravity model as
\begin{equation}\label{w1}
\omega_{R}
=-1-\frac{2}{3h}\left[\frac{\alpha X-1}{2X-1}\right],
\end{equation}
where
\begin{equation}\label{X}
X=\frac{\beta(h+1)^{\alpha-2}}{6c^2}\Big(\frac{R}{6h(2h+1)}\Big)^{\frac{\alpha-2}{2}}
=\frac{\beta(h+1)^{\alpha-2}}{6c^2}\Big({\frac{1}{1+z}}\Big)^{\frac{\alpha-2}{h}},
\end{equation}
and $z =\frac{1}{a}-1$ is the redshift. We take $a_0=1$ for the
present value of the scale factor.

For $\alpha<0,~\beta<0$ and $-1<h,~(h\neq0)$ one can find that the
EoS parameter (\ref{w1}) can cross the phantom-divide line.

For the second class of scale factors, the resulting $f(R)$ is
\begin{eqnarray}
f(R)=\lambda_+R^{m_+}+\lambda_-R^{m_-}-\frac{c^2(1-h)^2R}{h^2}+
~~~~~~~~~~\nonumber\\
\frac{4\beta(2h-1)(1-h)^\alpha R^{\frac{\alpha}{2}}}
{\big(h(\alpha-4)+(\alpha-1)(\alpha-2)\big)\big(6h(2h-1)\big)^{\frac{\alpha}{2}}},\label{fR-PLECHDE2}
\end{eqnarray}
where
\begin{equation}\label{alpha2}
m_\pm=\frac{3-h\pm\sqrt{h^2+10h+1}}{4}.
\end{equation}
\section{PLECNA $f(R)$-gravity model}
The new agegraphic dark energy (NADE) density is given by \cite{wei}
\begin{equation}\label{NADE}
    \rho_\Lambda=\frac{3n^2M_p^2}{\eta^2},
\end{equation}
where $\eta$ is the conformal time defined as
\begin{equation}\label{eta}
    \eta=\int^a_0\frac{dt}{Ha^2}.
\end{equation}
The power-law entropy corrected version of the NADE density takes
the form \cite{shay}
\begin{equation}\label{PLECADE}
 \rho_{\Lambda}=M_p^2\Big(\frac{3n^2}{\eta^2}-\frac{\beta}{\eta^{\alpha}}\Big).
\end{equation}
For the first class of scale factors (\ref{a}) and using Eq.
(\ref{respect to r}), the conformal time $\eta$ yields
\begin{equation}\label{eta2}
\eta=\int_t^{t_s}\frac{{\rm d}t}{a}
=\frac{(t_s-t)^{h+1}}{a_0(h+1)}
=\frac{1}{a_0(h+1)}\Big(\frac{6h(2h+1)}{R}\Big)^{\frac{h+1}{2}}.
\end{equation}
Note that due to having $\eta>0$ then $-1<h$.

Substituting Eq. (\ref{eta2}) into (\ref{PLECADE}) one can get
\begin{equation}\label{p2}
\rho_{\Lambda}=M_p^2\left[\left(\frac{3a_0^2n^2(h+1)^2}{\big(6h(2h+1)\big)
^{h+1}}\right)R^{h+1}-\left(\frac{\beta
a_0^2(h+1)^{2\alpha}}{\big(6h(2h+1)\big)
^{\frac{\alpha(h+1)}{2}}}\right)R^\frac{\alpha}{2}\right].
\end{equation}
Equating (\ref{density}) with (\ref{p2}) gives
\begin{eqnarray}\label{difeq3}
2R^2 f''(R)-(h+1)R f'(R)+(2h+1)f(R)+
~~~~~~~~~~~~~~~~~~~~~~~~~~~\nonumber\\
\left(\frac{a_0^2n^2(h+1)^2}{h\big(6h(2h+1)\big)
^{h}}\right)R^{h+1}-\left(\frac{2\beta
a_0^2(2h+1)(h+1)^{2\alpha}}{\big(6h(2h+1)\big)
^{\frac{\alpha(h+1)}{2}}}\right)R^\frac{\alpha(h+1)}{2}=0.
\end{eqnarray}
Solving Eq. (\ref{difeq3}) yields the PLECADE $f(R)$-gravity as
\begin{eqnarray}\label{fR-PLECADE}
f(R)=\lambda_+R^{m_+}+\lambda_-R^{m_-}-\frac{a_0^2n^2(h+1)^2R^{h+1}}{h^2(h+2)\big(6h(2h+1)\big)
^{h}}+
~~~~~~~~~~~~~~~~~~~~~~~~~~\nonumber\\
\frac{4\beta a_0^2(2h+1)(h+1)^{\alpha}R^\frac{\alpha(h+1)}{2}}{\big(6h(2h+1)\big)
^{\frac{\alpha(h+1)}{2}}\big((2-3\alpha+\alpha^2)+2h(2-2\alpha+\alpha^2)+\alpha (\alpha-1)h^2\big)},
\end{eqnarray}
where $m_\pm$ are given by (\ref{alpha}). Also $\lambda_\pm$ are
determined from the boundary conditions (\ref{bc1}) and (\ref{bc2})
as
\begin{eqnarray}\label{landa2}
    \lambda_\pm=\frac{1}{(m_\pm-m_\mp)R_0^{m\pm}}
    \Big[2m_\mp R_0+\frac{a_0^2n^2(h+1)^2(h+1-m_\mp)R_0^{h+1}}{h^2(h+2)\big(6h(2h+1)\big)^{h}}-
    ~\nonumber\\
    \frac{2\beta a_0^2(2h+1)(h+1)^{2\alpha}\big(\alpha(h+1)-2m_\mp\big) R_0^{\frac{\alpha(h+1)}{2}}}
    {\big(2(2h+1)-(h+1)(h+3)\alpha+(h+1)^2\alpha^2\big)\big(6h(2h+1)\big)^{\frac{\alpha(h+1)}{2}}}\Big].
\end{eqnarray}
Replacing Eq. (\ref{fR-PLECADE}) into (\ref{wHDETotal}) and using
(\ref{respect to r}) and (\ref{eta2}) one can get the EoS parameter
of the PLECNADE $f(R)$-gravity model as
\begin{equation}\label{w2}
\omega_{R}
=-1-\frac{2(h+1)}{3h}\left[\frac{\alpha X-1}{2X-1}\right],
\end{equation}
where
\begin{equation}\label{X2}
X=\frac{a_0^{(\alpha-2)}\beta(h+1)^{(\alpha-2)}}{6n^2}\Big(\frac{R}{6h(2h+1)}\Big)^{\frac{(\alpha-2)(h+1)}{2}}
=\frac{\beta(h+1)^{(\alpha-2)}}{6n^2}\Big(\frac{1}{1+z}\Big)^{\frac{(\alpha-2)(h+1)}{h}}.
\end{equation}
Equation (\ref{w2}) shows that for $\alpha<0,~\beta<0$ and
$-1<h,~(h\neq0)$ crossing the phantom-divide line can occur.

For the second class of scale factors we find
\begin{eqnarray}\label{fR-PLECADE2}
f(R)=\lambda_+R^{m_+}+\lambda_-R^{m_-}+\frac{a_0^2n^2(h-1)^2\big(6h(2h-1)\big)
^{h}R^{1-h}}{h^2(h-2)}-
~~~~~~~~~~~~~~~~\nonumber\\
\frac{4\beta a_0^2(2h-1)(h-1)^{2\alpha}R^\frac{\alpha(1-h)}{2}}{\big(6h(2h-1)\big)
^{\frac{\alpha(1-h)}{2}}\big((2-3\alpha+\alpha^2)-2h(2-2\alpha+\alpha^2)+\alpha (\alpha-1)h^2\big)},
\end{eqnarray}
where $m_\pm$ are given by Eq. (\ref{alpha2}).

\section{$f(R)$ reconstruction in de Sitter space}
The scale factor in de sitter space is defined as
\begin{equation}
a(t)=a_0e^{Ht},~~~H={\rm constant},\label{infaltion a}
\end{equation}
which can describe the early-time inflation of the universe
\cite{NojiriOdin2006}. Using Eqs. (\ref{R}) and (\ref{infaltion a})
one can obtain
\begin{equation}
H=\Big(\frac{R}{12}\Big)^{1/2}.\label{H inf}
\end{equation}
Then Eqs. (\ref{density}) and (\ref{pressure}) take the forms
\begin{equation}
\begin{array}{l}
k^2\rho_{R} = -\frac{1}{2}f(R)+3
H^2f'(R),\label{density i}\\
k^2p_{R}= \frac{1}{2}f(R)-3H^2f'(R).
\end{array}
\end{equation}
Also the EoS parameter yields $\omega_R=\frac{p_{R}}{\rho_{R}}=-1$
which behaves like the cosmological constant.

\subsection{ PLECHDE model}
For the scale factor (\ref{infaltion a}), using Eq. (\ref{H inf})
the future event horizon $R_h$ reduces to
\begin{equation}
R_h=a\int_t^{\infty}\frac{{\rm
d}t}{a}=H^{-1}=\left(\frac{R}{12}\right)^{-1/2}.\label{RhQ}
\end{equation}
Substituting Eq. (\ref{RhQ}) into PLECHDE density (\ref{PLECHDE})
yields
\begin{equation}\label{Poly Equ i}
R f'(R) -2 f(R)-c^2R+4\beta\Big(\frac{R}{12}\Big)^{\frac{\alpha}{2}}
=0.
\end{equation}
This gives
\begin{equation}\label{Poly f i}
f(R)=\lambda R^2-c^2R-\frac{8\beta}{\alpha-4}\Big(\frac{R}{12}\Big)^{\frac{\alpha}{2}}.
\end{equation}
where $\lambda$ is an integration constant. Also from
$\omega_R=\frac{p_{R}}{\rho_{R}}=-1$ and continuity equation for
PLECHDE one can get
\begin{equation}\label{beta}
    \beta=\frac{c^2R}{2\alpha}\Big(\frac{12}{R}\Big)^{\frac{\alpha}{2}}.
\end{equation}
Finally one can rewrite (\ref{Poly f i}) as
\begin{equation}\label{Poly f f}
f(R)=\lambda R^2-\frac{c^2(\alpha-2)^2}{\alpha(\alpha-4)}R.
\end{equation}
Note that the term $R^2$ confirms that the model (\ref{Poly f f})
satisfies the inflation condition \cite{Starobinsky}.

\subsection{PLECADE model}
For PLECADE, the conformal time $\eta$ for the scale factor (\ref{infaltion a}) yields
\begin{equation}
\eta=\int_0^t\frac{{\rm
d}t}{a}=\frac{1}{a_0H}\Big(1-e^{-Ht}\Big).\label{etadS1}
\end{equation}
Here to obtain $\eta=\eta(R)$ one cannot replace $t$ by $R$ in
(\ref{etadS1}). Therefore for the scale factor (\ref{infaltion a})
one cannot obtain the $f(R)$-gravity models corresponding to the
PLECNADE density (\ref{PLECADE}). To avoid of this problem we set
$t\rightarrow\infty$ for the upper limit of the integral
(\ref{etadS1}). Hence the result yields
\begin{equation}
\eta=\int_0^{\infty}\frac{{\rm
d}t}{a}=(a_0H)^{-1}=\left(\frac{a_0^2R}{12}\right)^{-1/2}.\label{etadS}
\end{equation}
Substituting Eq. (\ref{etadS}) into PLECNADE density (\ref{PLECADE})
and using Eq. (\ref{density i}) one can obtain
\begin{equation}\label{Poly Equ ii}
R f'(R) -2
f(R)-a_0^2n^2R+4\beta\Big(\frac{a_0^2R}{12}\Big)^{\frac{\alpha}{2}}
=0.
\end{equation}
Solving the above differential equation yields
\begin{equation}\label{Poly f ii}
f(R)=\lambda
R^2-a_0^2n^2R-\frac{8\beta}{\alpha-4}\Big(\frac{a_0^2R}{12}\Big)^{\frac{\alpha}{2}},
\end{equation}
where $\lambda$ is an integration constant. Also from
$\omega_R=\frac{p_{R}}{\rho_{R}}=-1$ and continuity equation for
PLECNADE one can get
\begin{equation}\label{beta2}
    \beta=\frac{a_0^2n^2R}{2\alpha}\Big(\frac{12}{a_0^2R}\Big)^{\frac{\alpha}{2}},
\end{equation}
and we finally find
\begin{equation}\label{Poly f f2}
f(R)=\lambda R^2-\frac{a_0^2n^2(\alpha-2)^2}{\alpha(\alpha-4)}R.
\end{equation}
Here like the model (\ref{Poly f f}) the inflation condition is
respected because of the term $R^2$.
\section{Conclusions}
Here, we investigated the power-law entropy corrected versions of
the HDE and NADE in the framework of $f(R)$ modified gravity. We
reconstructed different $f(R)$-gravity models corresponding to the
PLECHDE and PLECNADE models. We obtained the EoS parameters of the
PLECH and PLECNA $f(R)$-gravity models. Our calculations show that
for the selected models, the transition from the quintessence state
to the phantom regime can occur which is compatible with the recent
observations. Furthermore, we studied the PLECH and PLECNA
$f(R)$-gravity models in de Sitter space. We concluded that these
models can also satisfy the inflation condition.
\\
\\
\noindent{{\bf Acknowledgements}}\\
The work of K. Karami has been supported financially by Research
Institute for Astronomy $\&$ Astrophysics of Maragha (RIAAM),
Maragha, Iran.
\\


\end{document}